\documentclass[12pt]{article}
\begin{document}
\begin{titlepage}
\setlength{\textwidth}{5.9in}
\begin{flushright}
EFI 2000-59\\
 
\end{flushright}
\begin{center}
\vskip 1.0truein
{\Large\bf {Superconvergence Relations}}
\footnote{For the {\it `Concise Encyclopedia of SUPERSYMMETRY'}, \\
Kluwer Academic Publishers, Dortrecht, (Editors: Jon Bagger, Steven Duplij
and Warren Siegel) 2001.}

\vskip0.4truein
{Reinhard Oehme}
\footnote{E-mail: oehme@theory.uchicago.edu}
\vskip0.4truein
{\it Enrico Fermi Institute and Department of Physics}\\
{\it University of Chicago} \\
{\it Chicago, Illinois, 60637, USA}
\end{center}

\vskip1.0truein
  
\centerline{\bf Abstract}

\vskip0.2truein

For a limited number of matter fields, the discontinuity
of the transverse gauge field propagator can satisfy an
exact sum rule. With controlled and limited gauge dependence,
this superconvergence relation is of physical interest.
\end{titlepage}

\newpage
\baselineskip 18 pt
\pagestyle{plain}
\setlength{\textwidth}{5.9in}

SUPERCONVERGENCE RELATIONS are sum rules for the discontinuity
$\rho(k^{2})$ of a structure function $D(k^{2})$ in gauge
theories [1]. The transverse gauge field propagator is considered
here as a characteristic example. From Lorentz covarriance and a minimal
spectral condition, it follows that $D(k^{2})$ is the boundary
value of a function which is holomorphic in the complex $k^{2}$
plane with a cut along the positive real axis. A priori, with
the space-time propagator being a tempered distribution, this
function is bounded by a polynomial for $k^{2}\rightarrow\infty$.
But for regions in the parameter space where the theory
is asymptotically free, the asymptotic forms can be
calculated in terms of the weak coupling limit. Using
renormalization group methods, one finds for the structure function

\begin{eqnarray}
-k^2 D(k^2,\kappa^2,g,\alpha) & \simeq& \frac{\alpha}{\alpha_0} +
C(g^2,\alpha) \left(-\beta_0 \ln \frac{k^2}
{\kappa^2}\right)^{-\gamma_{00}/\beta_0} + \cdots ~,
\end{eqnarray}

\noindent and the asymptotic term for the discontinuity along the
positive, real $k^2$-axis is given by

\begin{eqnarray}
-k^2 \rho (k^2,\kappa^2,g,\alpha) & \simeq& \frac{\gamma_{0
0}}{\beta_0} C (g^2,\alpha) \left(-\beta_0 \ln \frac{k^2}{\vert
\kappa^2\vert}\right)^{-\gamma_{0 0}/\beta_0 - 1} + \cdots ~.
\end{eqnarray}

\noindent Here $\kappa^2 < 0 $ is the normalization point, and

\begin{eqnarray}
\gamma(g^2,\alpha) & ~=~ &(\gamma_{00} + \alpha \gamma_{01} ) g^2 ~+~
\cdots, \\
\beta(g^2) & ~=~ &\beta_0 g^4 ~+~ \cdots 
\end{eqnarray} 

\noindent are the limits
$g^2 \rightarrow 0$ of the anomalous dimension and the
renormalization group function, while $\alpha_0 = - {\gamma_{00} }/
{\gamma_{01}}$. It is important that the functional form of the
asymptotic discontinuity is independent of the
gauge parameter $\alpha \geq 0 $.

\noindent Given the analytic and asymptotic properties of  $D(k^2)$,
this function satisfies the dispersion relation

\begin{eqnarray}
D(k^2) = \int _{-0}^{\infty}d{k'}^2 \frac{\rho({k'}^2)}{({k'}^2-k^2)}~.
\end{eqnarray}

\noindent For the special case
$\gamma_{00}/\beta_0>0$, there is sufficient boundedness for the
validity of the Superconvergence Relation [1]

\begin{eqnarray}
\int_{-0}^{\infty}d k^2 \rho (k^2, \kappa^2, g , \alpha)
{}~~= ~~\frac{\alpha}{\alpha_0} ~,
\end{eqnarray}

\noindent which is most directly useful in the Landau gauge,
where $\alpha=0$:

\begin{eqnarray}
\int_{-0}^{\infty}d k^2 \rho (k^2, \kappa^2, g , 0)
{}~~= ~~0 ~.
\end{eqnarray}

The superconvergence relations are exact, and are not valid
order by order in perturbation theory. They
depend upon short- and long-distance properties of the theory.
Given $\beta_0 < 0$, it is the sign of  the anomalous dimension
coefficient $\gamma_{00}$ which determines the existence or
non-existence of superconvergence in all gauges.
In connection with the BRST cohomology, the superconvergence
relations can be used in order
to argue for the absence of the transverse gauge field
quanta from the physical state space of the theory (confinement) [2].
Hence $\gamma_{00}
=0$ indicates a phase transition similar to $\beta_0=0$. 

\noindent For SQCD with
the gauge group $SU(N_C)$ and $N_F$ flavors, the zero of
$\gamma_{00}(N_F,N_C)$ is at $N_F=\frac{3}{2}N_C$, and that of
$\beta_{0}(N_F,N_C)$ at $N_F=3N_C$. These values of 
$N_F$ define the lower and the upper limit of the conformal window [3].
The superconvergence relations are valid for $N_F<\frac{3}{2}N_C$.
As described above, for these values of $N_F$, the gauge quanta are
not in the physical state space.

\noindent There are also superconvergence relations for 
non-supersymmetric theories, and results similar to those
for SQCD were obtained earlier for QCD [2]. With the same gauge group 
$SU(N_C)$ and $N_F$ flavors, one has
superconvergence and confinement for $N_F<\frac{13}{4}N_C$. The window is
given by  $\frac{13}{4}N_C<N_F<\frac{22}{4}N_C$.

It is of interest to consider the possibility of superconvergence
relations for theories with space-time non-commutativity. When time
is involved, the integrand of a possible sum rule would be expected 
to have contributions from discontinuities of tachyonic branch cuts 
which are not directly associated with the input field content 
of the theory [4]. Dispersion relations for scattering amplitudes
are also modified by additional singularities assocated with 
violations of locality [5]. 

\noindent The new singularities are very different from the anomalous 
thretholds, which can be present even in local field theories.
These thresholds are well understood as structure singularities 
associated with the composite structure of the particles involved in
the theory [6]. If followed into secondary Riemann sheets by changing
the mass variables, the anomalous thretholds are seen to be ordinary
crossed-channel thretholds (left hand cuts) of amplitudes connected
by unitarity to the one under consideration. There are no structure
singularities in the gauge field propagator. 

\noindent With appropriately generalized BRST-methods, possible 
superconvergence relations could be of interest for the interpretation
of non-commutative gauge theories.

\vskip1.0truein

\newpage

BIBLIOGRAPHY.

[1] R. Oehme, W. Zimmermann, Phys. Rev. {\bf D21} (1980) 471;
                             Phys. Rev. {\bf D21} (1980) 1661;
    R. Oehme, Phys. Lett. {\bf B252} (1990) 641;
    R. Oehme, W. Xu Phys. Lett. {\bf B333} (1994) 172, hep-th/9406081;
                                {\bf B384} (1994) 269, hep-th/9604021.

[2] R. Oehme, Phys. Rev. {\bf D42} (1990) 4209;
              Phys. Lett. {\bf B195} (1987) 60;
    K. Nishijima, Prog.Theor.Phys. {\bf 75} (1986) 1221.

[3] R. Oehme, `Superconvergence, Supersymmetry and
               Conformal Invariance', in
               Leite Lopes Festschrift, pp. 443-457,
               World Scientific 1988 (KEK Library Scan,
               HEP SPIRES SLAC: Oehme, R.);
    N. Seiberg, Nucl. Phys. {\bf B435} (1995) 129, hep-th/9411149;
    R. Oehme, Phys. Lett. {\bf B399} (1997) 67, hep-th/9701012;
             `Superconvergence and Duality', Concise Encylopedia of
              SUPERSYMMETRY, Kluver 2001, hep-th 0101021.  

[4] J. Gomis and T. Mehen, Nucl. Phys. {\bf B591} (2000) 265,
                           hep-th/0005129; 
    L. Alvarez-Gaume, J. Barb$\acute{o}$n and R. Zwicky, hep-th/0103069;
                      (these papers contain further references). 
 
[5] R. Oehme,  `Forward Dispersion Relations and Microcausality',
              in Quanta, Wentzel Festschrift (University of Chicago
              Press, 1970) pp. 309-337; Phys. Rev. {\bf 100} (1955) 1503;
              (these papers contain further references).             
               
[6] R. Oehme, `The Compound Structure of Elementary Particles', in
              Werner Heisenberg und die Physik unserer Zeit, 
              Festschrift (Verlag Friedrich Vieweg und Sohn, 
              Braunschweig, 1961) pp. 240-259;
              Phys. Rev. {\bf 111} (1958) 1430; 
              Phys. Rev. {\bf 121} (1961) 1840;
              Nuovo Cimento {\bf 13} (1959) 778;
              (these papers contain further references).

\end{document}